\documentclass{llncs}

\usepackage{epsfig,url}
\usepackage{warpcol,xspace,subfigure}

\newtheorem{thrm}{Theorem}

\def\SA{\mbox{\rm SA}}

\def\ndoc{\mbox{\rm {\em ndoc}}}

\newcommand{\Oh}[1]
    {\ensuremath{\mathcal{O} \hspace{-.5ex} \left( {#1} \right)}}
\newcommand{\rank}[2]
    {\ensuremath{\mathrm{rank}_{#1} \hspace{-.5ex} \left( {#2} \right)}}

\def\+{\!+\!}
\def\-{\!-\!}

\def\pref(#1,#2){$#1$ is a prefix of $#2$}
\def\suff(#1,#2){$#1$ is a suffix of $#2$}
\def\reg(#1,#2){$#2$ is $#1$-regular}
\def\notreg(#1,#2){$#2$ is not $#1$-regular}

\def\eqref#1{(\ref{#1})}

\newlength{\onedigit}
\newlength{\onecomma}
\begin{document}

\title{\bf Range Quantile Queries:\\ Another Virtue of Wavelet Trees
  \thanks{This work was supported by the Sofja Kovalevskaja Award from the
Alexander von Humboldt Foundation and the German Federal Ministry of
Education and Research and by the Australian Research Council.}
}

\author{
Travis Gagie\inst{1}
\and
Simon J. Puglisi\inst{2}\thanks{Corresponding Author.}
\and
Andrew Turpin\inst{2}
}

\institute{
    Research Group for Combinatorial Algorithms in Bioinformatics,\\
    Bielefeld University, Germany\\
    \email{travis.gagie@gmail.com}\\[1ex]
 \and
    School of Computer Science and Information Technology,\\
    Royal Melbourne Institute of Technology, Australia\\
    \email{\{simon.puglisi,andrew.turpin\}@rmit.edu.au}
}


\maketitle \thispagestyle{empty}

\begin{abstract}
We show how to use a balanced wavelet tree as a data structure that stores
a list of numbers and supports efficient {\em range quantile queries}. A range
quantile query takes a rank and the endpoints of a sublist and returns the
number with that rank in that sublist.  For example, if the rank is half the
sublist's length, then the query returns the sublist's median. We also show
how these queries can be used to support space-efficient {\em coloured range
reporting} and {\em document listing}.
\end{abstract}

\section{Introduction}

If we are given a list of the closing prices of a stock for the past $n$
days and asked to find the $k$th lowest price, then we can do so in $\Oh{n}$
time~\cite{BFP+73}.  We can also preprocess the list in $\Oh{n \log n}$ time
and store it in $\Oh{n}$ words such that, given $k$ later, we can find the answer
in $\Oh{1}$ time: we simply sort the list.  However, we might also later face
{\em range quantile queries}, which have the form ``what was the $k$th lowest
price in the interval between the $\ell$th and the $r$th days?''.  Of course,
we could precompute the answers to all such queries, but storing them would
take \(\Omega (n^3 \log n)\) bits of space. In this paper we show how to use a
balanced wavelet tree to store the list in $\Oh{n}$ words such that we can
answer range quantile queries in $\Oh{\log \sigma}$ time, where $\sigma$ is the
number of distinct items in the entire list.
We can generalize our result to any constant number of dimensions but, currently, only by using slightly super-linear space.

We know of no previous work on quantile queries\footnote{Henceforth, for brevity,
we will use ``quantile query'' to mean ``range quantile query'', and similarly
with other types of range queries.}, but several authors have
written about {\em range median queries}, the special case in which $k$ is half the
length of the interval between $\ell$ and $r$.  Krizanc, Morin and Smid~\cite{KMS05}
introduced the problem of preprocessing for median queries and gave four
solutions, three of which have worse bounds than using a balanced wavelet tree;
their fourth solution involves storing $\Oh{n^2 \log \log n / \log n}$ words to
answer queries in $\Oh{1}$ time. Bose, Kranakis, Morin and Tang~\cite{BKMT05}
then considered approximate queries, and Har-Peled and Muthukrishnan~\cite{HM08}
and Gfeller and Sanders~\cite{GS??} considered batched queries. Recently, Krizanc
{\em et al.}'s fourth solution was superseded by one due to Petersen and
Grabowski~\cite{Pet08,PG09}, who reduced the space bound to
$\Oh{n^2 (\log \log n)^2 / \log^2 n}$ words. Table~\ref{tab:prev} shows the bounds for
Krizanc {\em et al.}'s first three solutions, for Petersen and Grabowski's solution,
and for using a balanced wavelet tree.

Har-Peled and Muthukrishnan~\cite{HM08} describe applications of median
queries to the analysis of Web advertising logs. In the final section of this
paper we show that our solution for quantile queries can be used to support
{\em coloured range reporting}, that is, to enumerate the distinct items in a
sublist. This result immediately improves V{\"a}lim{\"a}ki and M{\"a}kinen's recent
space-efficient solution to the {\em document listing problem}~\cite{m2002,vm2007}.

In the full version of this paper we will also discuss how to use a wavelet tree to
answer range counting queries (see~\cite{MN07}), coloured range counting queries
(returning the number of distinct elements in a range without enumerating them), and
how to support updates at the cost of slowing queries down to take time proportional
to the logarithm of the largest number allowed.

\begin{table}[t]
\begin{center}
\caption{Bounds for range median queries.}
\label{tab:prev}
\begin{tabular}{l|@{\hspace{2ex}}l@{\hspace{2ex}}l@{\hspace{2ex}}l}
& space (words) & time & restriction\\
\hline\\[-2ex]
Krizanc {\em et al.}~\cite{KMS05} & $\Oh{n}$ & $\Oh{n^\epsilon}$ & \(\epsilon > 0\)\\[.5ex]
Krizanc {\em et al.}~\cite{KMS05} & $\Oh{n \log_b n}$ & $\Oh{b \log^2 n / \log b}$ & \(2 \leq b \leq n\)\\[.5ex]
Krizanc {\em et al.}~\cite{KMS05} & $\Oh{n \log^2 n / \log \log n}$ & $\Oh{\log n}$ &\\[.5ex]
Petersen and &&&\\[-1.75ex]
    & $\Oh{n^2 (\log \log n)^2 / \log^2 n}$ & $\Oh{1}$ &\\[-1.75ex]
Grabowski~\cite{PG09} &&&\\[.5ex]
Theorem~\ref{thrm-quantile} & $\Oh{n}$ & $\Oh{\log n}$ &
\end{tabular}
\end{center}
\end{table}

\section{Wavelet Trees}

Grossi, Gupta and Vitter~\cite{GGV03} introduced wavelet trees for use in data
compression, and Ferragina, Giancarlo and
Manzini~\cite{FGM06} showed they have myriad virtues in this respect. Wavelet
trees are also important for compressed full-text indexing~\cite{nm2007}.
As we shall see, there is yet more to this intriguing data structure.


A wavelet tree $T$ for a sequence $s$ of length $n$ is an ordered, strictly
binary tree whose leaves are labelled with the distinct elements in $s$ in
order from left to right and
whose internal nodes store binary strings. The binary string at the root
contains $n$ bits and each is set to 0 or 1 depending on whether the corresponding
character of $s$ is the label of a leaf in $T$'s left or right subtree. For each
internal node $v$ of $T$, the subtree $T_v$ rooted at $v$ is itself a wavelet tree
for the {\em subsequence} of $s$ consisting of the occurrences of its leaves' labels.
For example, if \(s = \mathsf{a, b, r, a, c, a, d, a, b, r, a}\) and the leaves in
$T$'s left subtree are labelled {\sf a}, {\sf b} and {\sf c}, then the root
stores \(00100010010\), the left subtree is a wavelet tree for {\sf abacaaba} and
the right subtree is a wavelet tree for {\sf rdr}.
The important properties of the wavelet tree for our purposes are summarized
in the following lemma.
\begin{thrm}[Grossi et al.~\cite{GGV03}]
\label{thrm-wave-preprocess}
The wavelet tree $T$ for a list of $n$ elements on alphabet $\sigma$ requires
$n\log\sigma(1 + o(1))$ bits of space, and can be constructed in $O(n\log \sigma)$ time.
\end{thrm}

To see why the space bound is true, consider that the binary strings' total length
is the sum over the distinct elements of their frequencies times their depths, which
is $\Oh{n \log \sigma}$ bits. The construction time bound is easy to see from the
recursive description of the wavelet tree given above.

We note as an aside that, while investigating
data structures that support rank and select queries, M\"{a}kinen and Navarro~\cite{MN07}
pointed out a connection between wavelet trees and a data structure due to
Chazelle~\cite{Cha88} for two-dimensional range searching on sets of points.

\section{Range Quantile Queries}

We now describe how the wavelet tree can be used to answer quantile queries.
Let $s$ be the list of $n$ numbers we want to query. We build and store the wavelet
tree $T$ for $s$ and, at each internal node $v$, we store a small data structure that
lets us perform $\Oh{1}$-time rank queries on $v$'s binary string. A rank query on a
binary string takes a position and returns the number of 1s in the prefix that ends
at that position. Jacobson~\cite{Jac89} and later Clark~\cite{c1996} showed we can
support $\Oh{1}$-time rank queries on a binary string with a data structure that uses
a sublinear number of extra bits, beyond those needed to store the string itself. It
follows that the size of this preprocessed wavelet tree remains
$\Oh{n\log \sigma}$ bits.

\begin{figure}[t]
\begin{center}
\begin{tabular}{l@{\hspace{15ex}}l}
\includegraphics[width=60ex]{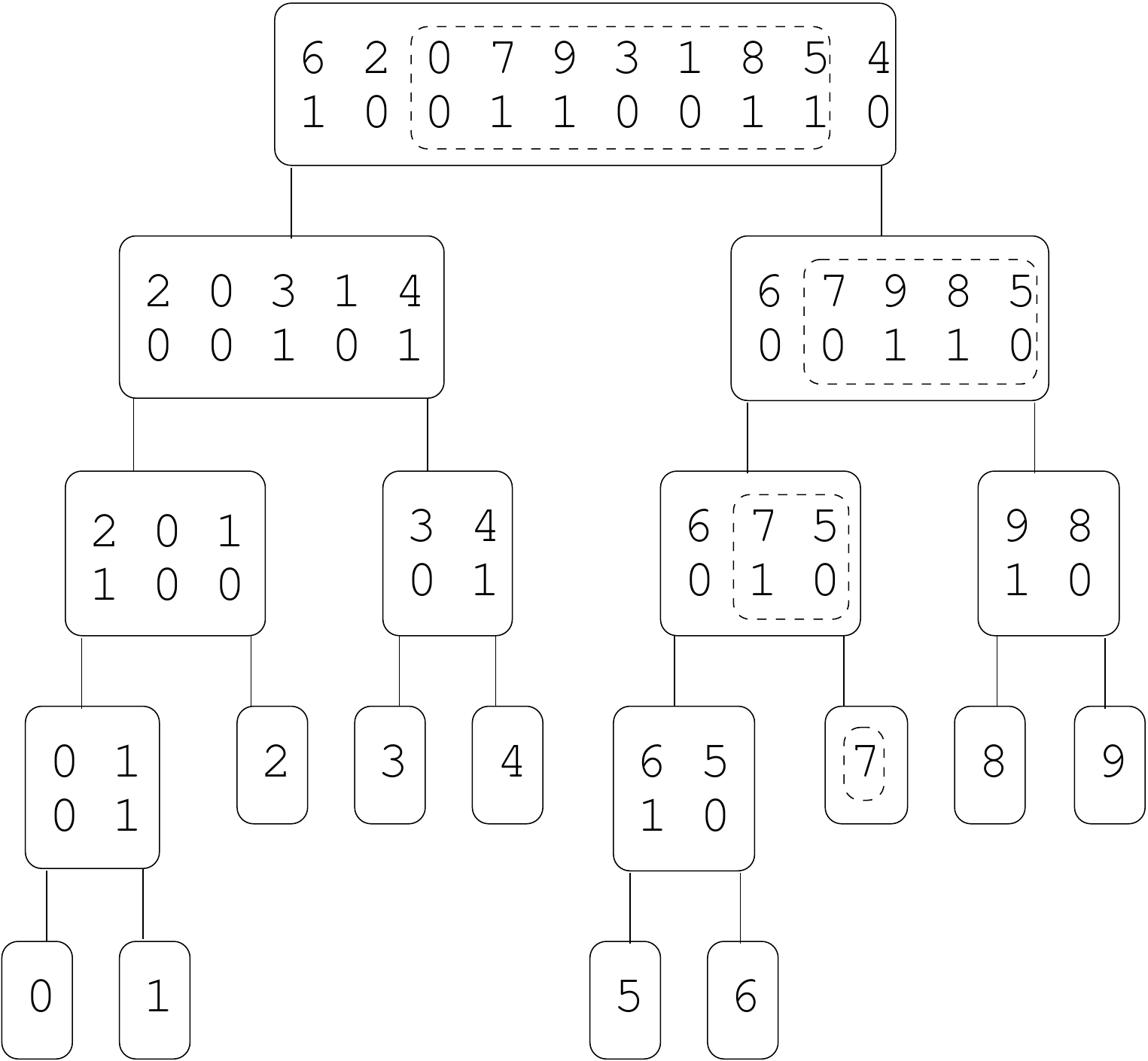} &
\raisebox{32ex}
{\(\begin{array}{l}
k = 5\\
\ell = 3\\
r = 9\\[3ex]
k = 2\\
\ell = 2\\
r = 5\\[3ex]
k = 2\\
\ell = 2\\
r = 3\\[3ex]
k = 1\\
\ell = 1\\
r = 1
\end{array}\)}
\end{tabular}
\end{center}
\caption{A wavelet tree $T$ (left) for \(s = 6, 2, 0, 7, 9, 3, 1, 8, 5, 4\), and the values (right) the variables $k$, $\ell$ and $r$ take on as we search for the 5th smallest element in \(s [3..9]\).  The dashed boxes in $T$ show the ranges from which we recursively select.}
\label{fig:example}
\vspace{2ex}
\end{figure}

Given $k$, $\ell$ and $r$ and asked to find the $k$th smallest number in \(s [\ell..r]\),
we start at the root of $T$ and consider its binary string $b$.  We use the two rank
queries $\rank{b}{\ell - 1}$ and $\rank{b}{r}$ to find the numbers of 0s and 1s
in \(b [1..\ell - 1]\) and \(b [\ell..r]\).  If there are more than $k$ copies of
0 in \(b [\ell..r]\), then our target is a label on one of the leaves in $T$'s left
subtree, so we set $\ell$ to one more than the number of 0s in \(b [1..\ell - 1]\),
set $r$ to the number of 0s in \(b [1..r]\), and recurse on the left subtree. Otherwise,
our target is a label on one of the leaves in $T$'s right subtree, so we subtract from
$k$ the number of 0s in \(b [\ell..r]\), set $\ell$ to one more than the number of 1s
in \(b [1..\ell - 1]\), set $r$ to the number of 1s in \(b [1..r]\), and recurse on the
right subtree.  When we reach a leaf, we return its label.
An example is given in Figure~\ref{fig:example}.
Since $T$ is balanced and we
spend constant time at each node as we descend (using the rank structures), our search
takes $\Oh{\log \sigma}$ time. Thus, together with Theorem~\ref{thrm-wave-preprocess}
we have the following.

\begin{thrm}
\label{thrm-quantile}
There exists a data structure of size $\Oh{n\log \sigma}$ bits which can be
built in $\Oh{n\log\sigma}$ time that answers range quantile queries on $s[1..n]$
in $\Oh{\log \sigma}$ time.
\end{thrm}

Some comments on $\sigma$ are in order at this point.
Firstly, and obviously, if $\sigma$ is constant, then so is our query
time.
If we represent the binary strings at each level of the
wavelet tree with a more complicated rank/select data structure of
Raman et.
al~\cite{rrr2002} (instead of Clark~\cite{c1996},
see~\cite{GGV03,MN07}), the size of the wavelet tree is reduced to
$nH_0(s) + \Oh{n\log\log n/\log_{\sigma} n}$ bits without affecting the
query time, where $H_0(s)$ is the zeroth order entropy of $s$. Prior solutions for
median queries do not make such {\em opportunistic} use of space.

At the other extreme, if $\sigma$ is $\Omega(n)$ we can map
the symbols in $s$ to the range $[1..n]$,
by first sorting the items in $\Oh{n\log n}$ time,
and storing the mapping in $\Oh{n\log\sigma}$ bits of space.
Preprocessing the array this way, and then using the wavelet tree
approach above, allows us to match the $\Omega(n\log n)$ time lower bound for
median queries~\cite{KMS05}, when the number of queries is $\Oh{n}$.
This lower bound applies to any computational
model which has an $\Omega(n\log n)$ time lower bound on sorting $s$.
Still, the solution is not completely satisfying, and we leave an open
question: Does an $\Oh{n\log n}$ preprocessing algorithm exist that
allows quantile (or even just median) queries to be answered in $o(\log
n)$ time when $\sigma$ is $\Omega(n)$?

It is not difficult to generalize Theorem~\ref{thrm-quantile} to any constant number of dimensions, using slightly super-linear space.  Suppose we are given a given a multidimensional array $A$ of total size $N$.  We build a balanced binary search tree on the $\sigma'$ distinct elements in $A$ and, at each node $v$, we store a binary array of size $N$ with 1s indicating the positions of occurrences of elements in $v$'s subtree.  We store each binary array in a folklore data structure (see, e.g.,~\cite[Lemma 2]{KN09}) that supports multidimensional range counting in $\Oh{1}$ time using $\Oh{m N^\epsilon}$ bits, where $m$ is the number of 1s and $\epsilon$ is any positive constant; thus, we use a total of $\Oh{N^{1 + \epsilon} \log \sigma'}$ bits.  To find the $k$th smallest number in a given range in $A$, we start at the root of the tree and use a range counting query to find the numbers of 0s and 1s in the same range of the binary array stored there.  If there are more than $k$ copies of 0 in the range, then we recurse on the left subtree; otherwise, we subtract the number of 0s from $k$ and recurse on the right subtree.  Since we use a single range counting query at each node as we descend, we use a total of $\Oh{\log \sigma'}$ time.

\begin{theorem}
\label{thm:multidimensional}
For any constants $d$ and \(\epsilon > 0\), there exists a data structure of size $\Oh{N^{1 + \epsilon} \log \sigma'}$ bits that answers $d$-dimensional range quantile queries on $A$ in $\Oh{\log \sigma'}$ time.
\end{theorem}

\section{Application to Space Efficient Document Listing}

The algorithm for quantile queries just described can, when coupled with
another wavelet tree property, be used to enumerate
the $d$ distinct items in a given sublist $s[\ell..r]$ in $\Oh{d\log\sigma}$
time as follows. Let $c_1, c_2, \ldots, c_d$ be the distinct elements in
$s[\ell..r]$ and, without loss of generality, assume $c_1 < c_2 < \ldots < c_d$.
Further, let $m_i$, $i \in 1..d$ be the number of times $c_i$ occurs in $s[\ell..r]$.
To enumerate the $c_i$, we begin by finding $c_1$, which can be achieved in
$\Oh{\log\sigma}$ via a quantile query, as $c_1$ must be the element with
rank~$1$ in $s[\ell..r]$. Observe now that $c_2$ must be the element in the
range with rank $m_1+1$, and in general $c_i$ is the element with rank $1+\sum_{j=1}^{i-1}{m_{j+1}}$.
Fortunately, each $m_i$ can be determined in $\Oh{\log\sigma}$ time by exploiting
a well known property of wavelet trees, namely, their ability to return,
in $\Oh{\log\sigma}$ the number of occurrences of a symbol in a prefix of
$s$~(see \cite{GGV03}). Each $m_i$ is the difference of two such queries.

The {\em document listing problem}~\cite{m2002} is a variation on the classical
pattern matching problem. Instead of returning all the positions at which a pattern
$P$ occurs in the text $T$, we consider $T$ as a collection of $k$ documents
(concatenated) and our task is to return the set of documents in which $P$ occurs.

Muthukrishnan~\cite{m2002}, who first considered the problem, gave an $\Oh{n\log n}$
bit data structure (essentially a heavily preprocessed suffix tree) that lists documents
in optimal $\Oh{|P| + \ndoc}$ time, where $\ndoc$ is the number of documents containing $P$.
Recently, V{\"a}lim{\"a}ki and M{\"a}kinen~\cite{vm2007} used more modern compressed and
succinct data structures to reduce the space requirements of Muthukrishnan's approach at
the cost of slightly increasing search to $\Oh{|P| + \ndoc \log k}$ time. Their data structure
consists of three pieces: the {\em compressed suffix array} (CSA) of $T$; a wavelet tree
built on an auxilliary array, $E$ (described shortly); and a succinct range minimum query
data structure~\cite{f2007}.

Central to both Muthukrishnan's and V{\"a}lim{\"a}ki and M{\"a}kinen's solutions is
the so-called ``document array'' $E[1..n]$, which is parallel to the suffix array $\SA[1..n]$:
$E[i]$ is the document in which suffix $\SA[i]$ begins. Given an interval $\SA[i..j]$
where all the occurrences of a pattern lie, the document listing problem then
reduces to enumerating the distinct items in $E[i..j]$.
Without getting into too many details, V{\"a}lim{\"a}ki and M{\"a}kinen use the
{\em compressed suffix array} (CSA) of $T$ to find the relevant sublist of $E$ in
$\Oh{|P|}$ time, and then a combination of $E$'s wavelet tree and a range minimum
query data structure~\cite{f2007} to enumerate the distinct items in that sublist
in $\Oh{\ndoc \log k}$ time. However, as we have described above, the wavelet tree of $E$
alone is sufficient to solve this problem in the same $\Oh{\ndoc \log k}$ time bound. In
practice we may expect this new approach to be faster, as the avoidance of the minimum
queries should reduce CPU cache misses. Also, because the wavelet tree of $E$ is already
present in~\cite{vm2007} we have reduced the size of their data structure by $2n + o(n)$
bits, the size of the data structure for minimum queries.

\section*{Acknowledgements}
Our thanks go to the three anonymous reveiwers whose helpful comments materially improved the paper,
and to Meg Gagie for righting our grammar.

\bibliographystyle{splncs_srt}
\bibliography{wavelet}

\end{document}